\documentclass[a4paper,pra,preprint,floatfix,superscriptaddress]{revtex4}
\usepackage[english]{babel}
\usepackage{amsmath,amsfonts,amssymb,float}
\usepackage{graphicx}
\usepackage{dcolumn,bm}
\usepackage{verbatim} 
\usepackage{color}
\usepackage{url}
\usepackage{float}
\usepackage{tikz}
\usetikzlibrary{arrows.meta, math}

\def\clf{Central Laser Facility, STFC Rutherford Appleton Laboratory, Didcot, OX
11 0QX, United Kingdom}
\def\strathclyde{Department of Physics, SUPA, University of Strathclyde, Glasgow, G4 0NG, United Kingdom}

\def\cockcroft{The Cockcroft Institute, Sci-Tech Daresbury, Warrington WA4 4AD, United Kingdom}

\def\Z{\mathbb{Z}}

\def\R{\mathbb{R}}

\renewcommand{\vec}[1]{\ensuremath{\mathbf{#1}}}

\begin{document}

\title{Dynamical symmetries in laser harmonic generation via a cubic nonlinearity}

\author{Raoul Trines}
\email[Corresponding author, ]{raoul.trines@stfc.ac.uk}
\author{Holger Schmitz}
\affiliation\clf
\author{Martin King}
\author{Paul McKenna}
\affiliation\strathclyde
\affiliation\cockcroft
\author{Robert Bingham}
\affiliation\clf
\affiliation\strathclyde

\date\today

\begin{abstract}
In our earlier work on harmonic generation with complex light [Nature Communications {\bf 15}, 6878 (2024)], we demonstrated how the harmonic spectrum of a complex laser beam in a nonlinear medium can be obtained through the judicious application of the ``beat wave'' concept. In this paper, we show how the same results can be obtained via the full set of symmetries of the initial laser-target configuration, and how this can be reconciled with the ``beat wave'' approach. We also highlight the connections between our work and existing theory for diffraction of EM waves from crystals: Laue equations, Mathieu equation, and theorems by Noether, Floquet and Bloch. The specific nature of our approach to harmonic spectra allows these connections to be revealed. We illustrate this with numerous examples taken from existing literature to show the wide applicability of our approach.
\end{abstract}

\maketitle

\section{Introduction}

In harmonic generation in nonlinear media by laser beams with complex configurations and/or topologies, selection rules are vital to determine the composition of the harmonic modes. However, there does not seem to be much order in the various ways that selection rules are derived, and there is even some controversy regarding what a selection rule actually is \cite{fleischer14, pisanty14, milo15}. In this paper, we aim to bring some order and clarity to the derivation of harmonic selection rules.

Selection rules for high harmonic generation are often derived via photon counting methods in the light of various conservation laws, in particular for the case of HHG in laser-gas interaction \cite{bertrand,gariepy,zli,kong,kong19,fleischer14,pisanty14,milo15,paufler}. A second approach, in particular for the case of HHG in laser-crystal interactions \cite{alon98,bayku,saito17, symmetry1, symmetry2, symmetry3, symmetry4, symmetry5}, is to derive the harmonic progression via the symmetries of the governing equations. In this paper, we build on our earlier work \cite{trines24} to show how both approaches can be reconciled. We also show how previous controversies \cite{fleischer14,pisanty14,milo15} can be reconciled and how our work complements and enhances earlier results \cite{symmetry1, symmetry3, symmetry4, symmetry5}.

We enhance our approach further to show how our results can be connected to earlier results on spectral selection rules based on symmetries, such as Noether's Theorem \cite{noether} and the Laue equations \cite{laue, kittel}. Overlap with theorems by Mathieu, Floquet and Bloch \cite{mathieu, floquet, bloch} is also found.

\section{Definitions}
\label{sec:defs}

We start from the real-valued vector potential of an EM wave with frequency $\omega$, wave number $k_z$, orbital angular momentum (OAM) level $\ell$ and some level of elliptical polarisation $\epsilon$: $\vec{A} \propto (\cos(\phi), \epsilon\sin(\phi))$ with $\phi \equiv \omega t - k_z z - \ell\varphi$, $|\epsilon| \leq 1$ and $\omega \geq 0$ always. We define $\vec{E} = -\partial\vec{A}/\partial t$, $\vec{B} = \nabla\times \vec{A}$, $\mathcal{E} = \epsilon_0 \vec{E}^2$, $P_z = \epsilon_0 (\vec{E} \times \vec{B})_z$, $L_z = \epsilon_0 \vec{E} \cdot (\vec{r}\times\nabla) \vec{A}$ and $S_z = \epsilon_0 (\vec{E}\times\vec{A})_z$. We also define the photon number density $N=\mathcal{E}/(\hbar\omega)$. Then we find that $\mathcal{E} = \hbar\omega N$, $P_z = \hbar k_z N$ and $L_z = \hbar\ell N$ as functions of $t$, $z$ and $\varphi$ are all proportional to $N\propto 1 + [(1-\epsilon^2)/(1+\epsilon^2)]\cos(2\phi)$, while $S_z = \epsilon$ does not depend on either $t$, $z$ or $\varphi$. We find that $S_z = \hbar\sigma N$ only when $\epsilon^2 = 1$, which removes the fluctuating part of $N$. Thus, the only polarisations that are eigenfunctions for $\mathcal{E}$, $P_z$, $L_z$ and $S_z$ at the same time are the pure circular polarisations with $\epsilon = \pm 1$. (We note that if a nonzero function $\Psi$ is a simultaneous eigenfunction of two operators $O_1$ and $O_2$ with nonzero eigenvalues $\lambda_1$ and $\lambda_2$, then we must have $\langle \Psi | O_1 | \Psi \rangle /\lambda_1 = \langle \Psi | O_2 | \Psi \rangle /\lambda_2 = \langle \Psi | \Psi \rangle$. Thus, a function $\Psi$ can only be a simultaneous eigenfunction for both $\mathcal{E}$ and $S_z$ if these two quantities have the same shape, which is only the case for $\epsilon^2 = 1$.) Since we wish to account for the conservation of energy, momentum and spin simultaneously, the pure circular polarisations (CP) are the ones we are compelled to use throughout this paper. For a more formal treatment including the Wigner function, see Mendon\c{c}a \cite{titobook}.

We also note that the pure CP mode has constant $|\vec{A}|^2$, meaning that it does not ``beat'' and has the highest possible symmetry. Thus, it cannot generate any harmonics in an isotropic medium. Adding a second CP mode causes $|\vec{A}|^2$ to beat and thus lowers the symmetry of the configuration, allowing harmonic generation. Since two pure CP modes beat together in precisely one well-defined way, each additional independent CP mode breaks the symmetry of the configuration in one specific way and allows harmonic generation in one independent direction in Fourier space. Linearly polarised light already has non-constant $|\vec{A}|^2$, and is thus less fundamental for our purposes than CP light. These are further arguments for the use of pure CP modes to study harmonic generation.

Next, we consider that the spin $\sigma = \pm 1$ of an EM wave is a physical quantity with a physical dimension. Thus, expressions like $\vec{A} = [\cos(\omega t), \sigma\sin(\omega t)]$ (one component is multiplied by spin, the other is not) or $\sigma^2 = 1$ (left-hand side is a physical quantity, right-hand side is a dimensionless constant) do not really make physical sense, even if they work mathematically, and are therefore to be avoided. Furthermore, we require that for the vector potential of a CP mode that is an eigenfunction for both energy and spin, that $\mathcal{E} = (\omega/\sigma) S_z$. Within these constraints, the only viable expression for such a vector potential is $\vec{A} = (\cos(\omega t/\sigma), \sin(\omega t/\sigma))$. This motivates our use of the quantities $\omega/\sigma$, $\vec{k}/\sigma$, $\ell/\sigma$ throughout this paper. Also note that this helps to distinguish between the strictly positive energy $\hbar\omega$ and the signed angular speed $\omega/\sigma$. Using $\omega/\sigma$ has the added advantage that two CP modes A and B always beat with the difference ``frequency'' $\omega_A/\sigma_A - \omega_B/\sigma_B$, independent of the direction of circular polarisation of A and B. When using $\omega$ alone, the beat frequency can be either the difference or the sum frequency, which is more complicated.

For a pure CP mode travelling in the $z$-direction, we have (per photon) $\mathcal{E} = \hbar\omega$, $\vec{P} = \hbar\vec{k}$, $L_z = \hbar\ell$ and $S_z = \hbar\sigma$, with $\omega \geq 0$ always and $\sigma = \pm 1$. We then define the wave phase via $\hbar d\phi \equiv \mathcal{E} dt - \vec{P}\cdot d\vec{x}$ (which becomes $\hbar d\phi \equiv \mathcal{E} dt - P_z dz - L_z d\varphi$ in cylindrical coordinates) and the ``beat angle'' via $d\alpha \equiv (\hbar d\phi)/S_z$. Finally, for real-valued $A_{x,y}$, we define the complex potential $\Psi = A_x + iA_y = \cos(\alpha) + i\sin(\alpha) = \exp(i\alpha) = \exp [i(\omega t - \vec{k}\cdot\vec{x} )/\sigma]$, or $\Psi = \exp [i(\omega t - k_z z - \ell \varphi )/\sigma]$ in cylindrical coordinates. These conventions are a formalisation of those used in our earlier work \cite{trines24} and will be used throughout this paper. We note that $\partial_t \Psi = (i\omega/\sigma)\Psi$, so we can no longer assume that any eigenvalue of the operator $i\hbar \partial_t$ is always a positive-definite energy.

For future use, we also define the vectors $X \equiv [t, \vec{x}]$ and $K \equiv [\omega/\sigma, -\vec{k}/\sigma]$, and their inner product $K\cdot X \equiv (\omega t - \vec{k}\cdot\vec{x})/\sigma = \alpha$, so we can write $\Psi = \exp(i K\cdot X)$ and $\cos(\alpha_B - \alpha_A) = \cos[(K_B - K_A)\cdot X]$. Using this notation, for two modes A and B, we have $\Psi_A^* \Psi_B = \exp i(\alpha_B - \alpha_A) = \exp [i(K_B - K_A) \cdot X]$ and $\Psi_A^* \Psi_B + \Psi_B^* \Psi_A = 2\cos[ (K_B - K_A) \cdot X]$. Therefore, $|\Psi_A|^2$ is constant for a single fundamental mode, while $|\Psi_A + \Psi_B |^2 = |\Psi_A|^2 + |\Psi_B|^2 + \Psi_A^* \Psi_B + \Psi_B^* \Psi_A = |\Psi_A|^2 + |\Psi_B|^2 + 2\cos[ (K_B - K_A) \cdot X]$. We also note that for any translation operator $\mathrm{T:} X \to X + dX$ for fixed $dX$, the function $\Psi = \exp(i K\cdot X)$ is an eigenstate with eigenvalue $\lambda_\mathrm{T} = \exp(i K\cdot dX)$

\section{Harmonic progressions governed by even symmetries}

In this section, we will discuss the relationship between harmonic progressions and ``even'' symmetries (translations, rotations), characterised by $X \to X + dX$. ``Odd'' symmetries (reflections, inversions) will be discussed in Section \ref{sec:odd}.

\subsection{Preliminaries}

We start from a nonlinear wave equation for the EM vector potential $\vec{A}$ with a cubic nonlinear term:
\begin{equation}
\label{eq:1}
(\partial_t^2 - c^2 \nabla^2 + \Omega^2 + \Gamma^2 |\vec{A}|^2 ) \vec{A} = \vec{0}.
\end{equation}
We define a coordinate symmetry of equation (\ref{eq:1}) as any orthogonal coordinate transformation that maps solutions onto other solutions. This requires that such transformations leave the operator $(\partial_t^2 - c^2 \nabla^2 + \Omega^2 + \Gamma^2 |\vec{A}|^2 )$ invariant. In practice, this means that an allowed symmetry must leave $|\vec{A}|^2$ invariant, while $\vec{A}$ is allowed to have a phase rotation. The reasons we study the cubic nonlinearity is that (i) it is a very common nonlinearity for harmonic generation, especially in isotropic media like a noble gas, and (ii) any symmetries that leave $|\vec{A}|^2$ invariant will also leave $\mathcal{E}$, $\vec{P}$, $L_z$ and $S_z$ invariant, and can thus be used with common conservation laws. As noted in Section \ref{sec:defs}, a pure CP mode has constant $|\vec{A}|^2$ and ``infinite'' symmetry with respect to $|\vec{A}|^2$ and can thus be considered a ``fundamental'' solution to Eq. (\ref{eq:1}).

Equation (\ref{eq:1}) admits several simple extensions. A second-order nonlinear term can be incorporated via the introduction of a ``DC mode'' and ``completing the square'': $|\vec{A}|^2 + 2\vec{A}\cdot \vec{A}_\mathrm{DC} = |\vec{A} + \vec{A}_\mathrm{DC}|^2 - |\vec{A}_\mathrm{DC}|^2$ \cite{trines24}, where $\vec{A}_\mathrm{DC}$ plays the role of a zero-frequency pump mode. Also, a nonlinear factor of the form $|\vec{A}|^2 + 2\vec{A}\cdot \vec{r} + V(\varphi)$, where $V$ denotes a periodic potential with period $2\pi/N$ \cite{alon98, bayku}, can be represented using two DC modes with $\ell/\sigma = -1$ and $N-1$ respectively; see also the structured targets used in Ref. \cite{trines24}. This extends our work to the study of e.g. structured targets or crystals with rotational symmetry.

We emphasise that allowed symmetries must preserve $|\vec{A}|^2$, for the following reaons. In previous work, e.g. Lerner \emph{et al.} \cite{symmetry3}, symmetries preserving either $A_x^2 + A_y^2$ or $b^2 A_x^2 + A_y^2/b^2$, $b \not= 1$, are mixed up, as are fundamental modes with respect to these symmetries. This is problematic, since Lerner's symmetry $e_{n,m}$ does not preserve $|\vec{A}|^2  \vec{A}$ for $b \not= 1$, and thus does not map solutions of Eq. (\ref{eq:1}) onto other solutions. This leads to further confusion, as in Ref. \cite{fleischer14}, where a laser beam with elliptical polarisation (constant $b^2 A_x^2 + A_y^2/b^2$ for $b \not= 1$, not isotropic) is used as a fundamental mode for HHG in a noble gas (isotropic, so $A_x^2 + A_y^2$ should have been conserved). In response, several papers stated that the pure CP modes should be used as fundamental modes, but without clearly stating why \cite{pisanty14, milo15}. We stress that the proper procedure is: (i) determine which quantity should be preserved under symmetries, as determined by the model under consideration, e.g. $A_x^2 + A_y^2$ for Eq. (\ref{eq:1}); (ii) determine the fundamental modes that will preserve this quantity at all times, e.g. pure CP modes; (iii) decompose all EM fields into these fundamental modes and use those to study the harmonic generation process.

As in Section \ref{sec:defs} and in Ref. \cite{trines24}, we use $\Psi = A_x + iA_y$ and write $\Psi = \Psi_0 + \delta\Psi$, where $\Psi_0$ represents the driving laser beams and $\delta\Psi$ the harmonic modes. We write $\langle |\Psi_0|^2 \rangle$ for the constant average of the oscillating pump laser intensity. Inserting this into (\ref{eq:1}), we find:
\begin{align}
\label{eq:2}
(\partial_t^2 - c^2 \nabla^2 + \Omega^2 + \Gamma^2 |\Psi_0|^2 )\delta\Psi = -\Gamma^2 ( |\Psi_0|^2  - \langle |\Psi_0|^2 \rangle )\Psi_0,\\
\label{eq:2a}
(\partial_t^2 - c^2 \nabla^2 + \Omega^2 + \Gamma^2 \langle |\Psi_0|^2 \rangle ) \Psi_0 = -\Gamma^2 (\Psi_0^* \delta\Psi + \Psi_0 \delta\Psi^*)\Psi_0.
\end{align}
Here, Eq. (\ref{eq:2}) represents the growth of the harmonic modes, while Eq. (\ref{eq:2a}) represents the corresponding depletion of the pump pulse.
We multiply Eq. (\ref{eq:2})  by $\Psi_0^*$ to obtain:
\begin{equation}
\label{eq:rules}
\Psi_0^* (\partial_t^2 - c^2 \nabla^2 + \Omega^2 + \Gamma^2 |\Psi_0|^2 ) \delta\Psi = -\Gamma^2 ( |\Psi_0|^2  - \langle |\Psi_0|^2 \rangle )|\Psi_0|^2.
\end{equation}
From this, we obtain the following rules: (i) any \emph{allowed symmetry} will leave $\Psi_0^* \Psi_0$ invariant, and (ii) for any \emph{allowed harmonic} $\delta\Psi$, $\Psi_0^* \delta\Psi$ must remain invariant under any allowed symmetry. Note the correspondence with well-known theorems by Bloch and Floquet \cite{bloch, floquet}; see also Section \ref{sec:noether}.

\subsection{One-dimensional harmonic progression}
\label{sec:1-d}

For a single CP mode $\Psi_0 = \exp(i K_A \cdot X)$, we find that $\Psi_0^* \Psi_0 = 1$, so any even symmetry $X \to X + dX$ will preserve this constant function and is thus allowed. For a harmonic $\delta\Psi =  \exp(i K_n\cdot X)$, we find that $\Psi_0^* \delta\Psi =  \exp[i(K_n - K_A)\cdot X]$ must be preserved under any symmetry $X \to X + dX$,  so $\exp[i(K_n - K_A)\cdot dX] = 1$ for any $dX$. This implies that $K_n = K_A$, i.e. no harmonics are allowed in this case. This situation is depicted in Figure \ref{fig:1}(a).

Next, we consider two non-degenerate CP modes: $\Psi_0 = \exp(i K_A \cdot X) +  \exp(i K_B \cdot X)$ with $K_A \not= K_B$.  We use this $\Psi_0$ with Eq. (\ref{eq:rules}).  Since $\Psi_0^* \Psi_0 = 2 + 2\cos[(K_B - K_A)\cdot X]$, any allowed symmetry must satisfy $\cos[(K_B - K_A)\cdot (X + dX)] = \cos[(K_B - K_A)\cdot X]$, or $(K_B - K_A)\cdot dX = 2\pi m$, $m\in \Z$. 
We define $X_{AB} = 2\pi (K_B - K_A)/|K_B - K_A|^2$, so allowed symmetries satisfy $dX = mX_{AB} + \tilde{X}$ with $\tilde{X}$ such that $(K_B - K_A)\cdot \tilde{X} = 0$, i.e. a set of hyperplanes. For any allowed harmonic $\delta\Psi = \exp(i K_n\cdot X)$, $\Psi_A^* \delta\Psi$ must remain invariant under any allowed symmetry, i.e. $\exp[ i(K_n - K_A)\cdot dX] = 1$ for $dX = mX_{AB} + \tilde{X}$. Setting $m=0$ first, we find that $\exp[ i(K_n - K_A)\cdot \tilde{X}] = 1$ for any $\tilde{X}$ such that $(K_B - K_A)\cdot \tilde{X} = 0$. This implies that $K_n - K_A = \lambda(K_B -K_A)$ for some $\lambda$. For arbitrary $m$, we then find that $(K_n -K_A)\cdot dX = 2\pi m \lambda$, so $\exp[ i(K_n - K_A)\cdot dX] = 1$ implies $m \lambda = n \in \Z$. This leads to the general solution $K_n -K_A = n(K_B -K_A)$, $n \in \Z$, in agreement with Trines \emph{et al.} \cite{trines24}. Since $(K_B - K_A)\cdot \tilde{X} = 0$, we also find that $K_n \cdot \tilde{X} = K_A \cdot \tilde{X} = K_B \cdot \tilde{X}$, which defines a set of conservation laws. This situation is depicted in Figure \ref{fig:1}(b).

We also consider a configuration with a single pump mode $\Psi_0 = \exp(i K_A \cdot X)$ but subject to a single discrete symmetry, i.e. the configuration is invariant under $X \to X + X_1$ but not under $X \to X + \varepsilon X_1$ with $|\varepsilon| < 1$. This situation can arise e.g. for a laser beam hitting a crystal or other periodic structure. In similar fashion to the above, one can derive the harmonic progression $K_n -K_A = n K_1$, $n \in \Z$, with $K_1 = 2\pi X_1/|X_1|^2$. We will elaborate on this in the sections \ref{sec:multi} and \ref{sec:noether}.

Inserting $\Psi_0 = \exp(i K_A \cdot X) +  \exp(i K_B \cdot X)$ into Eq. (\ref{eq:2}) yields an equation similar to the driven Mathieu equation \cite{mathieu}, with a leading-order term nonlinear term $2\cos[(K_B - K_A)\cdot X] d\Psi$ and a driving term $(\Psi_0^* \Psi_0) \Psi_0$. Allowed symmetries of this equation also satisfy $(K_B - K_A)\cdot dX = 2\pi n$, $n\in \Z$, as illustrated in Figure 1a of Ref. \cite{kfir15}. While the spectrum of the homogenous Mathieu equation is centred around $0$, that of the driven Mathieu equation is centred around $K_{A,B}$, see also Section \ref{sec:noether}.

Expanding the expression $K_n = K_A + n(K_B - K_A)$ into its components yields the following harmonic progression $(n \in \Z)$ \cite{trines24}:
\begin{align}
\label{eq:spinenergy}
\omega_n/\sigma_n &= \omega_A/\sigma_A + n( \omega_B/\sigma_B - \omega_A/\sigma_A),\\
\vec{k}_{n}/\sigma_n &= \vec{k}_{A}/\sigma_A  + n( \vec{k}_{B}/\sigma_B - \vec{k}_{A}/\sigma_A ),\\
\label{eq:spinoam}
\ell_n/\sigma_n &= \ell_A/\sigma_A  + n( \ell_B/\sigma_B - \ell_A/\sigma_A ),\\
\label{eq:spinspin}
\sigma_n/\sigma_n &= \sigma_A/\sigma_A  + n(\sigma_B/\sigma_B - \sigma_A/\sigma_A ),
\end{align}
where Eq. (\ref{eq:spinenergy}) uniquely determines $\sigma_n$ and the last equation is trivially true, ensuring that every harmonic step is spin-neutral. If Eq. (\ref{eq:1}) has been obtained from a model that conserves energy, we can define $\#_{A,B}$ as the number of photons absorbed from channels A and B, and use energy conservation: $\omega_n = \#_A \omega_A + \#_B \omega_B$. This yields expressions for the photon numbers required to generate harmonic $n$ (via the lowest-order pathway): 
\begin{equation}
\label{eq:photonnumber}
(\#_A, \#_B)_n = (\sigma_n/\sigma_A, 0) + n(-\sigma_n/\sigma_A,  \sigma_n/\sigma_B).
\end{equation}
Using Eq. (\ref{eq:photonnumber}) with e.g. Eqns. (\ref{eq:spinoam}) or (\ref{eq:spinspin}) yields conservation laws for linear or angular momentum or spin.

\subsection{Multi-dimensional harmonic progression}
\label{sec:multi}

This procedure can be generalised to multi-dimensional space. Before we do so, we will introduce the concept of \emph{synthetic coordinates}. Assume one has a configuration of two LP pulses with unequal frequencies $\omega_{A,B}$. This leads to four fundamental modes with $\omega/\sigma = \pm \omega_A$ and $\pm \omega_B$. The harmonic progression is then given by $(\omega/\sigma)_{mn} = (2m+1)\omega_A + 2m\omega_B$. While the two indices seem to indicate a two-dimensional progression, the spectrum occupies only one dimension in Fourier space (frequency). This discrepancy can be resolved by introducing a synthetic coordinate $s$ and associated Fourier coordinate $k_s$. The four fundamental modes then become $(\omega/\sigma, k_s/\sigma) = \pm (\omega_A, k_A)$ and $\pm (\omega_B, k_B)$, and the harmonic progression becomes fully two-dimensional: $(\omega/\sigma, k_s/\sigma)_{mn} = (2m+1)(\omega_A, k_A) + 2m(\omega_B. k_B)$. When needed, one can remove the synthetic coordinate by setting $k_A = k_B = 0$, which recovers the original harmonic progression in $\omega/\sigma$.

The procedure to generate multi-dimensional progressions is then as follows.
\begin{enumerate}
\item Determine the full collection $\{dX\}$ of allowed even symmetries $X \rightarrow X + dX$ in $X$-space.
\item Determine a ``minimal cell'' in this collection (as known from crystal lattice theory \cite{kittel}), and determine a basis $\{X_i\}$ for this ``minimal cell''. If the cell is infinitesimal in one direction, denote the basis vector as $\epsilon X_i$ in that direction, where $\epsilon$ represents an infinitesimal step. Likewise, if the cell is unbounded in one direction, denote the basis vector as $\Lambda X_i$ in that direction, where $\Lambda = 1/\epsilon$ denotes an ``infinitely large'' step.
\item If ``synthetic'' coordinates were used to highlight the dimensionality of the configuration, then these should be included here also.
\item Determine a reciprocal basis $\{K_i\}$ such that $K_i\cdot X_j = 2\pi\delta_{ij}$. Since the vectors $\{X_i\}$ are independent, this is always possible. An infinitesimal (infinite) vector $\epsilon X_i$ ($\Lambda X_i$) will lead to an infinite (infinitesimal) reciprocal vector $\Lambda K_i$ ($\epsilon K_i$). Under a symmetry $X \to X + X_i$, two neighbouring modes in the direction $i$ develop a mutual phase difference of $2\pi$, while the phase difference between two neighbouring modes in any direction $j \not= i$ will not change.
\item Using a starting ``fundamental mode'' $ K_A$, e.g. from the driving laser beam, the harmonic progression now becomes:
  \begin{equation}
  \label{eq:kprogression}
  K_{\{n\}} - K_A  = \sum_i n_i K_i,
  \end{equation}
where all $n_i \in \Z$. This ensures that $(K_{\{n\}} - K_A)\cdot X_j = 2\pi n_j$ and $\exp [ i (K_{\{n\}} - K_A)\cdot X_j ] = 1$ for all $j$. This is equivalent to the set of Laue equations known from solid state physics \cite{kittel,laue}, but now extended to any finite number of dimensions.
\end{enumerate}

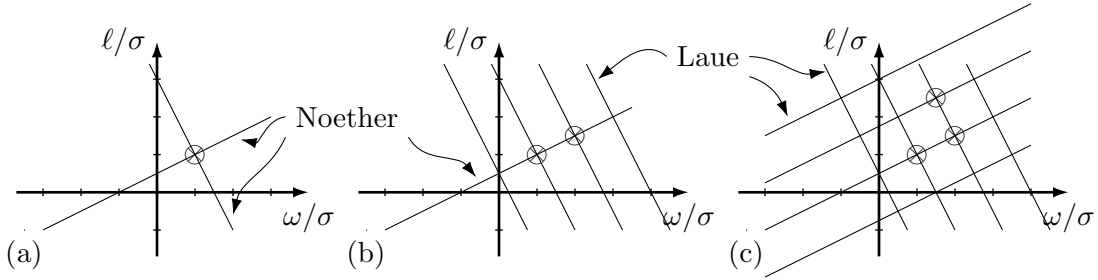
\begin{figure*}[ht]
\begin{tikzpicture}[scale=1]
    \begin{scope}[scale=0.5]
    \draw (-3.5,-1) -- (-3.5,-1) node[pos=1,below] {(a)};
      \draw[-{Latex[length=6pt, width=4pt]}, line width=1pt] (-3.7,0) -- (4,0) node[pos=1,below] {$\omega/\sigma$};
      \draw[-{Latex[length=6pt, width=4pt]}, line width=1pt] (0,-1.7) -- (0,4) node[pos=1,left] {$\ell/\sigma$};
      \foreach \x in {1,2,3} {
        \draw[line width=0.5pt] (\x,0.1) -- (\x,-0.1);
        \draw[line width=0.5pt] (-\x,0.1) -- (-\x,-0.1);
      }
      \foreach \y in {1,2,3}
        \draw[line width=0.5pt] (0.1,\y) -- (-0.1,\y);
      \draw[line width=0.5pt] (0.1,-1) -- (-0.1,-1);
    
      \draw[thin] (-0.2,3.4) -- (2,-1);   
      \draw[thin] (-3,-1) -- (3,2);       
    
      \node at (1,1) {$\otimes$};
    \end{scope}
    \begin{scope}[scale=0.5, xshift=9cm]
      \pgfmathsetmacro{\countA}{2}
      \pgfmathsetmacro{\posXA}{1}
      \pgfmathsetmacro{\posYA}{1}
      \pgfmathsetmacro{\incXA}{1}
      \pgfmathsetmacro{\incYA}{0.5}
      \pgfmathsetmacro{\incXB}{0.5}
      \pgfmathsetmacro{\incYB}{1.0}
      \pgfmathsetmacro{\hOffsetA}{\incXA + \incYA*\incXB/\incYB}
     \draw (-3.5,-1) -- (-3.5,-1) node[pos=1,below] {(b)};
      \draw[-{Latex[length=6pt, width=4pt]}, line width=1pt] (-3.7,0) -- (5,0) node[pos=1,below] {$\omega/\sigma$};
      \draw[-{Latex[length=6pt, width=4pt]}, line width=1pt] (0,-1.7) -- (0,4) node[pos=1,left] {$\ell/\sigma$};
      \foreach \x in {-3,...,4}
        \draw[line width=0.5pt] (\x,0.1) -- (\x,-0.1);
      \foreach \y in {1,2,3}
        \draw[line width=0.5pt] (0.1,\y) -- (-0.1,\y);
      \draw[line width=0.5pt] (0.1,-1) -- (-0.1,-1);
    
      \foreach \i in {-1,0,1,\countA}
        \draw[thin] (-0.2 + \i*\hOffsetA,3.4) -- (2 + \i*\hOffsetA, -1);   
      \draw[thin] (-3,-1) -- (3.5,2.25);       
    
      \foreach \i in {0,1}
        \node at (1 + \i, 1 + 0.5*\i) {$\otimes$};
    \end{scope}
    \begin{scope}[scale=0.5, xshift=19cm]
      \pgfmathsetmacro{\countA}{2}
      \pgfmathsetmacro{\countB}{2}
      \pgfmathsetmacro{\posXA}{1}
      \pgfmathsetmacro{\posYA}{1}
      \pgfmathsetmacro{\incXA}{1}
      \pgfmathsetmacro{\incYA}{0.5}
      \pgfmathsetmacro{\incXB}{-0.5}
      \pgfmathsetmacro{\incYB}{1.0}
      \pgfmathsetmacro{\hOffsetA}{\incXA - \incYA*\incXB/\incYB}
      \pgfmathsetmacro{\vOffsetB}{\incYB - \incXB*\incYA/\incXA}
      \draw (-3.5,-1) -- (-3.5,-1) node[pos=1,below] {(c)};
      \draw[-{Latex[length=6pt, width=4pt]}, line width=1pt] (-3.7,0) -- (5,0) node[pos=1,below] {$\omega/\sigma$};
      \draw[-{Latex[length=6pt, width=4pt]}, line width=1pt] (0,-1.7) -- (0,4) node[pos=1,left] {$\ell/\sigma$};
      \foreach \x in {-3,...,4} 
        \draw[line width=0.5pt] (\x,0.1) -- (\x,-0.1);
      \foreach \y in {-1,...,3}
        \draw[line width=0.5pt] (0.1,\y) -- (-0.1,\y);
    
      \foreach \i in {-1,0,1,\countA}
        \draw[thin] (-0.2 + \i*\hOffsetA, 3.4) -- (2 + \i*\hOffsetA, -1);   
      \foreach \i in {-1,0,1,\countB}
        \draw[thin] (-3,-1 + \i*\vOffsetB) -- (4, 2.5 + \i*\vOffsetB);       
    
      \foreach \i in {0,1}
        \node at (1 + \incXA*\i, 1 + \incYA*\i) {$\otimes$};
      \node at (1 + \incXA + \incXB, 1 + \incYA + \incYB) {$\otimes$};
    \end{scope}
    \node (noether) at (2.5,1) {Noether};
    \draw[-{Latex[length=6pt, width=4pt]}, bend left] (noether.west) to[out=340,in=140] (1.1,0.7);
    \draw[-{Latex[length=6pt, width=4pt]}, bend left] (noether.south west) to[out=10,in=200] (1.0,-0.3);
    \draw[-{Latex[length=6pt, width=4pt]}, bend left] (noether.south east) to[out=330,in=160] (4.2,0.2);
    \node (laue) at (7.25,1.8) {Laue};
    \draw[-{Latex[length=6pt, width=4pt]}, bend left] (laue.west) to[out=300,in=200] (5.8,1.5);
    \draw[-{Latex[length=6pt, width=4pt]}, bend left] (laue.east) to[out=10,in=200] (8.8,1.6);
    \draw[-{Latex[length=6pt, width=4pt]}, bend left] (laue.south east) to[out=30,in=160] (8.3,1.0);
\end{tikzpicture}

\caption{Hierarchy of harmonic progressions in 2-D $(\omega/\sigma, \ell/\sigma)$ space. (a) A single CP mode, $A^2$ obeys two independent continuous symmetries, given by $\epsilon X_1$ and $\epsilon X_2$, leading to two independent Noether-conserved quantities, indicated by two independent lines, given by $\Delta K \cdot X_1 = \Delta K \cdot X_2 = 0$. There is only one crossing, so no harmonics possible. (b) Two independent CP modes, one continuous symmetry of $A^2$ becomes discrete, ``is broken'', e.g. $\epsilon X_1$ becomes $X_1$. The ``broken'' symmetry leads to a collection of equally spaced parallel lines $\Delta K \cdot X_1 = 2\pi n_1$ (Laue), while the remaining continuous symmetry leads to a single line $\Delta K \cdot X_2 = 0$ (Noether). The line crossings lead to a 1-D harmonic progression. (c) Three independent CP modes, breaking the other continuous symmetry of $A^2$ as well. This leads to two sets of parallel lines $\Delta K \cdot X_1 = 2\pi n_1$ and $\Delta K \cdot X_2 = 2\pi n_2$; the crossings form a 2-D regular grid (Laue).}
\label{fig:1}
\end{figure*}

A number of consequences follows from the above procedure:
\begin{enumerate}
\item For a progression with 1 or 2 dimensions, we also note the similarity to the spectrum of the ``laser beat wave'' concept in plasma-based electron acceleration \cite{beatwave1,beatwave2,beatwave3,beatwave4,beatwave5,beatwave6,beatwave7}. Of these, Ref. \cite{beatwave2} explicitly couples the generation of beat waves to the Mathieu equation \cite{mathieu}. In similar fashion, the equations for Raman and Brillouin scattering by Forslund \emph{et al.} \cite{forslund} can also be rewritten to resemble the driven Mathieu equation.
\item For an infinite $\Lambda K_i$, the corresponding $n_i$ must be zero, since the result needs to be finite. For an infinitesimal $\epsilon K_i$, the implication is that the harmonic progression is continuous in that direction rather than discrete.
\item For a continuous symmetry, i.e. an infinitesimal $\epsilon X_i$ and an infinite $\Lambda K_i$, one finds necessarily that $n_i = 0$. Then we obtain an expression of the form $( K_{\{n\}} - K_A) \cdot X_i = 0$, so no harmonic progression in that direction in $K$-space. This same expression is also a conservation law. If all $X_i$ are infinitesimal, so all $K_i$ are infinite, then $n_i = 0$ for all $i$ and there will be no harmonic progression at all (see the case of a single CP wave in an isotropic medium given in Section \ref{sec:1-d}).
\item We note that the RHS of Eq. (\ref{eq:kprogression}) defines a hyperplane in $K$-space which includes the origin, so Eq. (\ref{eq:kprogression}) can  be used to derive many novel conservation laws similar to the ``conservation of torus-knot angular momentum'' \cite{torusknot}, a law which we prefer to write as $\tau\omega/\sigma - \ell/\sigma = \gamma$, or $\tau \Delta(\omega/\sigma) - \Delta (\ell/\sigma) =0$.
\item If some $X_i$ becomes finite (from infinitesimal), so one $K_i$ becomes finite also (from infinite), a harmonic progression will occur in the corresponding direction. This is the ``breaking'' of a continuous symmetry to become discrete. The more continuous symmetries are broken, the more harmonics there will be in the spectrum (and more dimensions to the harmonic progression).
\item In practice, symmetry breaking is often achieved via either the addition of one or two fundamental modes to an existing configuration \cite{symmetry4}, or via the interaction of the pump laser modes with a medium (crystal) with a discrete symmetry that is not already present in the pump laser (so the combination of laser plus medium has a lower symmetry than the laser by itself). This way, one creates a ``hierarchy'' of harmonic progressions, where $N+1$ independent pump modes (which may include a DC mode provided by a medium) drive an $N$-dimensional harmonic progression between them, and each additional independent pump mode breaks the symmetry in an additional independent way, and thus adds an independent dimension to the harmonic progression. This hierarchy is illustrated in Figure \ref{fig:1}; see also Section \ref{sec:noether}.
\end{enumerate}

The above procedure works well when using fundamental modes for the following reasons. (i) A single mode has constant $\Psi^* \Psi$ so does not change the set of allowed symmetries. (ii) Two modes A and B can beat in exactly one way, with the difference $K_B - K_A$, and will thus turn at most one continuous symmetry into a discrete one (symmetry breaking). This will lead to at most one new reciprocal vector $K_i$. (iii) As a consequence, when starting from a fundamental mode, each $K_i$ vector will lead to one new fundamental mode, and the harmonic progression of allowed fundamental modes can be reached via discrete $K_i$ steps.

In the case of elastic linear scattering of a single laser mode off a crystal, this procedure reduces to the Laue equations in solid-state physics \cite{laue,kittel}, which are restricted to the three spatial symmetries. We see that our procedure extends the Laue equations to spatiotemporal symmetries, i.e. augments them to 4-D. We will expand on this finding in Section \ref{sec:noether}. In certain cases where the pump contains many fundamental modes, one may wish to employ synthetic spatial dimensions to augment the Laue equations to 5 or more dimensions. This does not change the basic procedure though. Note that, as opposed to Ref. \cite{symmetry4}, we do not require synthetic dimensions in so-called ``tangent'' space, only in ``real'' $X$-space and Fourier $K$-space.

In addition to the overlap with the Laue and Mathieu equations \cite{laue,mathieu}, our approach also shows overlap with theorems by Noether \cite{noether}, Floquet \cite{floquet} and Bloch \cite{bloch}, as will be explained in Section \ref{sec:noether}. We can exploit this overlap because we consider symmetries that leave the Hamiltonian $A^2$ invariant, rather than the vector potential $\vec{A}$ or the nonlinear polarisation $A^2 \vec{A}$, and use ``fundamental'' modes that are also Floquet modes. This same overlap is not obvious in earlier approaches \cite{symmetry2,symmetry3,symmetry4,symmetry5}. In Ref. \cite{symmetry1}, the overlap with Floquet theory is obscured, since the authors use the symmetries from the standard Floquet group but do not match those symmetries to a specific Hamiltonian, while their harmonic model is not guaranteed to be a Floquet system. While some of the symmetries in Ref. \cite{symmetry1} preserve $A_x^2 + A_y^2$, others preserve $b^2 A_x^2 + A_y^2/b^2$ for $b \not= 1$; thus, it is never clear which modes are the pure Floquet modes.

For a configuration where all the fundamental modes are known, one can always calculate a collection $\{K_i\}$ from the difference vectors between the fundamental modes in $K$-space ($K_1 =  K_B -  K_A$, $K_2 =  K_C -  K_A$, etc.), adding ``synthetic'' dimensions where needed to turn the set $\{K_i\}$ into an independent basis. For a configuration where all the symmetries are known, one can use one fundamental mode from the pump laser and the collection $\{K_i\}$ to find all the fundamental modes needed to describe the configuration, adding additional fundamental modes if necessary (see also Ref. \cite{trines24}, where a DC mode is used with $\omega=0$ but a well-defined value for $\ell/\sigma$). Since the ``photon counting'' approach to HHG \cite{bertrand,gariepy,pisanty14,milo15,paufler} follows directly from the collection of fundamental modes, while the ``symmetry-based'' approach \cite{alon98,bayku,saito17,symmetry1,symmetry2,symmetry3} follows directly from the collection $\{K_i\}$, this actually unifies these two approaches to calculating the HHG selection rules.

For example, for the case of HHG from the interaction of a laser beam with a crystal \cite{alon98,bayku}, where the nonlinear term reads  $[\exp(i N\varphi) + \Psi\exp(-i\varphi)]\Psi$, we write $K_1 = (0,N)$, $K_2 =  K_A - (0,1)$, $K_3 =  K_B - (0,1)$, and so on. (The vector $K_1 = (0,N)$ follows from the $N$-fold rotational symmetry of the crystal. The more discrete symmetries the crystal has, the more such vectors there will be.) This way, the harmonic progression can be calculated much faster than when the symmetries are calculated first.

The above symmetry considerations can be extended from 2-D fields to 3-D fields, see e.g. Ref. \cite{symmetry1}, and also Section \ref{sec:3-d}. Like that work, we assume an EM field with a 3-D topology: there is a ``parallel'' field component $E_z$ in addition to $E_\perp$. Then $E_z^2 = E^2 - \Psi^* \Psi$, which contains terms $\cos( \theta_B/\sigma_B - \theta_A/\sigma_A)$ or similar. We consider a symmetry that leaves $\Psi^* \Psi$ invariant. If $E_z \to E_z$ under this symmetry, then we get integer harmonics of $ \theta_B/\sigma_B - \theta_A/ \sigma_A $. If $E_z \to -E_z$, then we get half-integer harmonics of $ \theta_B/\sigma_B - \theta_A/ \sigma_A$. Compare e.g. stimulated Raman scattering, where $E_z \propto \delta n \propto \Psi^* \Psi$ after linearization, so we always obtain integer harmonics of $ \theta_B/\sigma_B - \theta_A/ \sigma_A$.

\subsection{Counting photons and dimensions}

Equations (\ref{eq:photonnumber}) and.(\ref{eq:kprogression}) allow one to determine the number of dimensions needed to describe the harmonic progression and the number of photons from each fundamental mode or ``channel'' that are (minimally) needed to produce a given harmonic mode.
\begin{enumerate}
\item $N+1$ independent fundamental laser modes generate $N$ vectors $K_i$ between them, and thus an $N$-dimensional progression. Thus, the presence of three ``channels'' allows for two dimensions rather than three  \cite{pisanty14, milo15}. If one starts with $N$ vectors rather than fundamental modes, one should remember that the origin should be counted as point $N+1$.
\item If the initial fundamental modes are not fully independent, the dimension of the progression can be lower than $N$, but can be brought back up to $N$ by adding synthetic dimensions.
\item A laser beam with linear or elliptical polarisation counts as two modes, not one. Thus, one CP and one EP beam add up to three fundamental modes, and the resulting harmonic progression has two dimensions rather than one \cite{fleischer14}.
\item The $K$-vector provided by an even symmetry connects two fundamental modes. One needs at least one absolute mode from the pump laser. The symmetries then provide the other modes relative to the first absolute mode.
\item For HHG in laser-gas interaction, each vector $K_i$ connects two fundamental modes in $K$-space. For laser interaction with a crystal or a structured solid target, where a vector $K_i$ originates from a symmetry, it may be necessary to introduce ``DC'' fundamental modes for the purposes of photon counting, parity, conservation laws, etc. When energy, momentum or spin are exchanged with such a ``synthetic'' mode, this means in practice that these are absorbed by the target, see also Ref. \cite{trines24}.
\item A rotational crystal that is ``even'' under a minimal rotation provides the ``DC modes'' $(\omega/\sigma, \ell/\sigma) = (0,-1)$ and $(0,N-1)$. See e.g. the structured targets by Trines \emph{et al.} \cite{trines24}, for various $N$.
\item A rotational crystal that is ``odd'' under a minimal rotation provides the ``synthetic modes'' $(\omega/\sigma, \ell/\sigma) = (0,-N-1)$ and $(0,N-1)$. See e.g. the BBO crystal studied in Ref. \cite{tang}, with $N=3$.
\item In these cases, the rotational crystal provides two fundamental modes and at least one symmetry. The pump laser topology with respect to the crystal then provides other symmetries and the remaining points.
\item Once all fundamental modes and connecting $K$-vectors have been determined, one can  count all the photons and/or modes needed to generate a given harmonic mode. Spin and parity conservation both go via photon counting. If this is done correctly then both are always conserved.
\item While $\sigma_\mathrm{DC}$ is not defined for a DC mode, we note that $\sigma_\mathrm{DC} \#_\mathrm{DC}$ is well-defined for a DC mode in a given harmonic pathway, so DC modes can be used in e.g. Eqns. (\ref{eq:spinenergy}) and (\ref{eq:photonnumber}). We will expand on the ``proper'' counting of DC modes in future work.
\end{enumerate}

This further enables a discussion of the order of the shortest pathway to a given harmonic mode.
To start, the contribution of channel A to the order of a given harmonic is $|\#_A| = |\sigma_A \#_A|$. The full order of this harmonic is thus given by $\sum_i |\sigma_i \#_i|$. We note that, for two numbers $x, y \in \R$, the triangle inequality states that $||x| - |y|| \leq |x+y| \leq |x| + |y|$, and that $||x| - |y|| = |x+y|$ when $xy \leq 0$ while $|x+y| = |x| + |y|$ when $xy \geq 0$. For only two modes A and B, spin conservation implies $|\sigma_A \#_A  + \sigma_B \#_B| = 1$. Since $|\sigma_A \#_A|, |\sigma_B \#_B| \geq 1$, this means that $| |\sigma_A \#_A| - |\sigma_B \#_B|| = |\sigma_A \#_A  + \sigma_B \#_B|$, so $\sigma_A \#_A  \sigma_B \#_B \leq 0$ always. For a configuration with three or more modes, one can show that the terms $\sigma_i \#_i$ cannot all have the same sign, but for three modes A, B and C, this still allows $\sigma_A \#_A \sigma_B \#_B > 0$ as long as the sign of $\sigma_C \#_C < 0$ differs from the other two terms. For e.g. $\sigma_A \#_A > 0$, we have $|\sigma_A \#_A \pm 1| = |\sigma_A \#_A| \pm 1$, while for $\sigma_A \#_A < 0$, we have $|\sigma_A \#_A \pm 1| = |\sigma_A \#_A| \mp 1$. For $\sigma_A \#_A  \sigma_B \#_B > 0$ (only possible for 3 or more independent fundamental modes, i.e. a harmonic progression having 2 or more dimensions), we have $|\sigma_A \#_A \pm 1| + |\sigma_B \#_B \mp 1| = | \sigma_A \#_A| +  |\sigma_B \#_B |$ and a step (A,B) in the harmonic progression is order-neutral. For $\sigma_A \#_A  \sigma_B \#_B < 0$, we have $|\sigma_A \#_A \pm 1| + |\sigma_B \#_B \mp 1| = | \sigma_A \#_A| +  |\sigma_B \#_B | \pm 2$ and a step (A,B) in the harmonic progression changes the order of the harmonic by 2. This phenomenon can be seen clearly in the results of Ref. \cite{trines24b}, where the amplitudes of modes with $|k_\perp/k_0| < |\omega/\omega_0|$ are mostly independent of $|k_\perp|$, as a step in the $k_\perp$ direction is order-neutral. Meanwhile, the amplitudes of modes with $|k_\perp/k_0| > |\omega/\omega_0|$ decrease rapidly for increasing $|k_\perp|$, as the order changes by 2 for each step in the $k_\perp$ direction.

This discussion formalises various notions of ``perturbative OAM conservation rules'' put forward in earlier work \cite{rego16,rego19,zhang21}. It also contributes to the notion of the ``cutoff'' in non-perturbative harmonic generation. For HHG in the frequency domain (1-D progression), the cutoff is a single point \cite{lewenstein} or at best two points on the $\omega/\sigma$ axis. For HHG in the frequency-OAM domain (2-D progression), the cutoff was found to be two points on the OAM axis, whose location depends on the harmonic frequency \cite{rego16}. For a generic 2-D progression, the cutoff will become a closed curve in e.g. $(\omega/\sigma, \ell/\sigma)$ space. To determine the location and shape of that curve, it is important to determine the order of individual peaks in the 2-D progression also.

\subsection{Non-collinear 3-dimensional interaction of EM waves}
\label{sec:3-d}

So far, we have only considered laser modes that are travelling in (nearly) parallel directions, i.e. $k_\perp/k_\parallel \ll 1$. However, harmonic generation using non-collinear beams can occur in various scenarios. First, there is the configuration of so-called ``synthetic chiral light' consisting of non-collinear beams' that is used to diagnose left- and right-handed enantiomers of a certain chiral molecule \cite{ayuso}. Second, there is the generation of harmonics through nonlinear vacuum polarisation by two ultra-intense laser beams crossing at 90 degrees \cite{grismayer}. This has two consequences. First, the harmonic spectra will split in two subsets, each associated with the polarisation and propagation direction of one of the pump beams. Second, the contribution to $A^2$ of the beating between two non-collinear modes $\Psi_A$ and $\Psi_B$ becomes more complex than $\Re(\Psi_A^* \Psi_B)$.
For example, the interaction of two non-collinear linearly polarised pulses will have to be drawn in (at minimum) three-dimensional $(\omega/\sigma, k_x/\sigma, k_z/\sigma)$ space \cite{grismayer}.

Here, we consider two non-collinear beams A and B whose wave vectors lie in the $(x,z)$-plane, so the $y$-direction is the vertical direction for both. We define $x_A$ and $x_B$ as the ``in-plane'' transverse coordinate for each beam, and $\Psi_{A,B} = E_{xA,xB} + iE_y$, so $\Psi_{A,B}$ is defined with respect to the wave vector of the beam rather than a fixed laboratory coordinate system. As a first hypothesis, for a superposition of two fundamental (circularly polarised) modes A and B, we calculate:
\begin{align}
E^2 - c^2 B^2 &\propto 2  (\vec{A}_A \cdot \vec{A}_B) (\omega_A \omega_B - c^2 \vec{k}_A \cdot \vec{k}_B)/(\sigma_A \sigma_B),\\
\vec{A}_A \cdot \vec{A}_B &= (\Psi_A \Psi_B^* + \Psi_A^* \Psi_B) \cos(\alpha/2) + (\Psi_A \Psi_B + \Psi_A^* \Psi_B^*)\sin(\alpha/2)\\
\cos\alpha &= \vec{k}_A \cdot \vec{k}_B/(k_A k_B).
\end{align}
As a consequence, two circularly polarised modes intersecting at an angle $0 < \alpha < \pi$ generate extra terms in $\vec{A}_A \cdot \vec{A}_B$, so this behaves more like two linearly polarised modes. The additional terms lower the symmetry of the interaction and lead to the generation of a richer harmonic spectrum. Adding more fundamental modes (e.g. by changing the polarisations to linear), the number of interactions will increase, leading to an even richer harmonic spectrum. This is reflected by e.g. the larger number of 3-D symmetries discussed in Ref. \cite{symmetry1} compared to 2-D symmetries, and also by the ever more complex models developed to address fully 3-D fields \cite{symmetry5}.

\section{Connection with Noether's Theorem and spectral theorems}
\label{sec:noether}

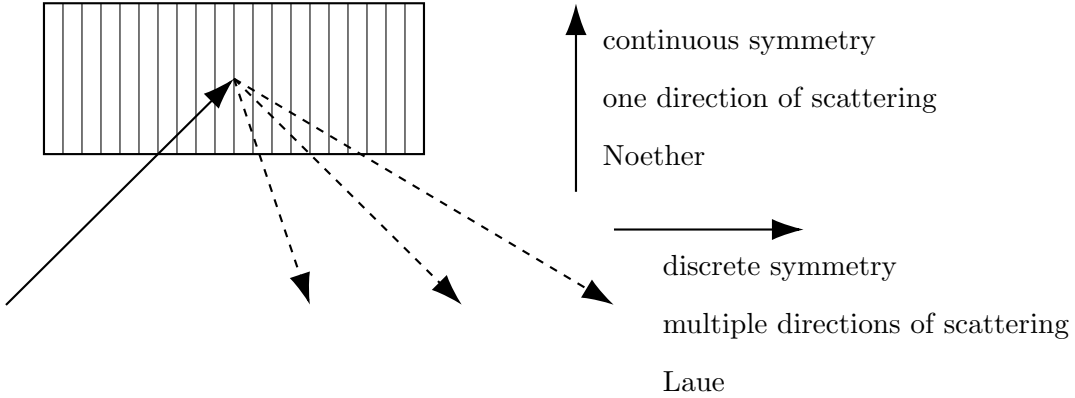
\begin{figure*}[ht]
\begin{tikzpicture}[scale=1.0]
\draw[thick] (5,0) rectangle (10,2);
\foreach \i in {1,...,19}
    \draw[thin] (5+0.25*\i,0) -- (5+0.25*\i,2);

\draw[-{Latex[length=12pt, width=8pt]}, thick] (4.5,-2) -- (7.5,1);
\draw[-{Latex[length=12pt, width=8pt]}, dashed, thick] (7.5,1) -- (10.5,-2);
\draw[-{Latex[length=12pt, width=8pt]}, dashed, thick] (7.5,1) -- (8.5,-2);
\draw[-{Latex[length=12pt, width=8pt]}, dashed, thick] (7.5,1) -- (12.5,-2);

\draw[-{Latex[length=12pt, width=8pt]}, thick] (12,-0.5) -- (12,2) node [below right=6pt, align=left] {
    continuous symmetry\\
    one direction of scattering\\
    Noether
};

\draw[-{Latex[length=12pt, width=8pt]}, thick] (12.5,-1) -- (15,-1)

node[below right, align=left] at (13,-1.2) {
    discrete symmetry\\
    multiple directions of scattering\\
    Laue
};
\end{tikzpicture}
\caption{A simple illustration of the connection between Noether's Theorem (vertical direction) and the Laue equations (horizontal direction) via light diffraction off a grating. See text for a full description.}
\label{fig:2}
\end{figure*}

In this section, we explore the connection between our work and Noether's Theorem \cite{noether}, and also with various spectral theorems by Laue \cite{laue}, Bloch \cite{bloch} and Floquet \cite{floquet}. The connection between Noether's and Laue's theorems is already suggested by the example of light diffraction from a grating with vertical grooves with horizontal spacing $L$, see Figure \ref{fig:2}. Shifting this grating by an infinitesimal vertical distance will not change anything, so the vertical $k$-component is the same for all diffraction orders (Noether). However, the grating needs to be shifted by a minimum distance of $L$ in the horizontal direction to map it back onto itself; no smaller shift will do. Thus, the horizontal $k$-components of the diffraction orders are separated by $2\pi/L$ (Laue). (When the minimal shift is ``infinitesimal'', then the separation between side bands becomes ``infinite'', i.e. there will be no side bands.) This shows how both Noether's and Laue's theorems can be present in the same configuration. The connection can also be seen in the ``hierarchy of progressions'' displayed in Figure \ref{fig:1}.

We demonstrate the connection between the two theorems for a periodic Lagrangian $L$. We take the following steps.
We start from a periodic Lagrangian, given by: $L \propto m_e c^2 \exp[ (i/\hbar) \int L' dt]$, with $L'$ of the form $L' dt = \mathcal{E} dt - \vec{p} \cdot d\vec{x}$ or similar.
We define with dimensionless ``action integrals'' $S(t) = \int (L/\hbar) dt$ and $S'(t) = \int (L'/\hbar) dt$, and also $\hbar K \equiv [\mathcal{E}, -\vec{p}]$. For the Hamiltonian $H$ and generalised momentum $\vec{P}$, we find $[H, -\vec{P}] \propto i\hbar [ \partial S /\partial t, \nabla S ] \propto  [\mathcal{E}, -\vec{p} ] S = (\hbar K) S$.
Of course, $L$ will be constant if $L'$ is constant. But if $\Delta \int L' dt = 2\pi \hbar$, then $L$ is also conserved, so $L$ may be conserved under discrete as well as continuous symmetries. This extension is important to make the connection between Noether's Theorem, the Laue equations and Eq. (\ref{eq:kprogression}) above.
We note that the derivative of $\exp(ix)$ is never zero for finite $x$, so if $L$ and $S$ are conserved under a continuous symmetry, then so is $S'$. This recovers Noether's original theorem: a continuous symmetry $X \to X + \varepsilon dX$ that leaves $S$ invariant leads to the conservation of $[ \partial S /\partial t, \nabla S ] \cdot dX$ and thus of $K \cdot dX$, i.e. $dK \cdot dX = 0$.
However, suppose $L$ is conserved under a discrete symmetry $X \to X + dX$ while it is not conserved under $X \to X + \varepsilon dX$ with $0 < |\varepsilon| < 1$. This allows $S' \to S' + 2\pi $ under this discrete symmetry, and in some cases it may even imply $S' \to S' + 2\pi \varepsilon$ under $X \to X + \varepsilon dX$ with $0 < |\varepsilon| < 1$.
In that case, the spectrum of S' satisfies $dK \cdot dX = 2\pi$ at minimum, often $dK \cdot dX = 2\pi n$, and the spectrum of $\exp(i S)$ definitely satisfies $dK \cdot dX =2\pi n$ via the theory of the driven Mathieu equation \cite{mathieu} and the Jacobi-Anger expansion.
We note the obvious overlap with the Laue equations \cite{laue,kittel} when the symmetries are purely spatial: $dX = [0; d\vec{x}]$.
Since $-i [ \partial L/\partial t , \nabla L] =  LK$ etc., and $[H, -\vec{P}] \propto (\hbar K) S$, one can thus imagine an extension to Noether's theorem, where $\exp[ i \Delta S'] = 1$ under a finite coordinate shift $X \to X + dX$, $dX = (dT, dX_i)$ (which is weaker than $\Delta S' = 0$), leading to expressions of the form $\hbar dK \cdot dX = (\Delta H) dT - (\Delta p_i) dX_i = 2\pi n\hbar$, $n \in\Z$, where $H$ and $p_i$ are taken from $L$ rather than $L'$. The case that $dX$ is infinitesimal requires $n=0$, and Noether's original theorem is recovered. This extension will also show overlap with the well-known Floquet and Bloch theorems \cite{bloch,floquet} from crystallography.

In the derivation above, we only provide the relative value $dK$. The starting point $K_0$ for the eventual harmonic progression is usually determined by a driving term, ``pump wave'', leading to expressions like $K_n \cdot X' = K_0 \cdot X'$ for a conserved quantity, or $K_n \cdot dX = K_0 \cdot dX + 2\pi n$ for a ``periodic'' quantity. Note that in the homogeneous Mathieu equation (without a driving term), the effective starting point is $K_0 = 0$.

It will thus be worthwhile to explore the following conjecture further: define $S \equiv \int L' dt$ and let $X \to X + dX$ be a discrete symmetry that leaves $L= \exp(i S/\hbar)$ invariant, while $\exp(i S/\hbar)$ is not invariant under $X \to X + \epsilon dX$ for any $|\epsilon| < 1$, then the ``Noether quantity'' $N' \equiv (N'_0,-N'_i) = [H', -P'] = \hbar K$ derived from $-i[\partial L/\partial t, -\nabla L] = LK$ satisfies $(dN')\cdot dX = 2\pi n \hbar$, $n\in\Z$, making use of the Jacobi-Anger expansion or equivalent. If we can define a basis $\{X_i\}$ for all the (discrete or continuous) symmetries of $\exp(i S'/\hbar)$, and a reciprocal basis $\{K_i\}$, then we may even propose the stronger conjecture that $dN' = \sum_i n_i \hbar K_i$. Since necessarily $n_i = 0$ for a continuous symmetry in the direction $X_i$, this will then default to $(dN)\cdot X_i = 0$ in that case. 


At this point, we recall that, for a pump wave with two fundamental modes, $|\Psi_A + \Psi_B|^2 = |\Psi_A|^2 + |\Psi_B|^2 + 2\cos[ (K_B - K_A) \cdot X]$. 
Furthermore, writing $\Psi = \Psi_0 + d\Psi = \Psi_A + \Psi_B + d\Psi$, we obtain (i) a nonlinear term given by $[ |\Psi_A|^2 + |\Psi_B|^2 + 2\cos[ (K_B - K_A) \cdot X] ] d\Psi$, clearly reminiscent of the Mathieu equation \cite{mathieu}, and (ii) a driving term $[ |\Psi_A|^2 + |\Psi_B|^2 + 2\cos[ (K_B - K_A) \cdot X] ] (\Psi_A + \Psi_B|)$, providing the starting points $K_A$, $K_B$ for the eventual harmonic progression.
In order to apply the above results for a periodic Lagrangian, we need to make the identification $ \exp[ (i/\hbar) \int L' dt] \leftrightarrow \cos[ (K_B - K_A) \cdot X]$. This can be done as follows: $ \int (L' / \hbar) dt \leftrightarrow \alpha_B - \alpha_A  = (K_B - K_A) \cdot X$, $dN' = [H', -P'] \leftrightarrow \hbar \Delta K = \hbar (K_n - K_A)$. For a continuous symmetry that conserves $\cos[ (K_B - K_A) \cdot X]$, we then find $(K_n - K_A) \cdot X = 0$, while for a discrete symmetry, $(K_n - K_A) \cdot X = 2\pi n$, see also Eq. (\ref{eq:kprogression}), demonstrating the overlap between our work and the extension to Noether's theorem.

We note that the conserved quantities in Noether's Theorem are usually $\mathcal{E}$ or $\vec{P}$, where these are $\mathcal{E}/S_z$ or $\vec{P}/S_z$ in our case; this follows directly from the nature of the cross terms $\Psi_A^* \Psi_B$, our definition of $d\alpha$ versus Noether's $Ldt$, and the fact that $\partial \Psi/\partial t \propto (\omega/\sigma)\Psi$ rather than $\propto \omega\Psi$. Thus, the connection of our work with Noether's Theorem for continuous symmetries and the Laue equations for discrete ones only becomes visible because we use ``fundamental modes'' with constant $\vec{A}^2$, and $(\omega/\sigma, \vec{k}/\sigma)$ rather than $(\omega, \vec{k})$, contrary to most other works on this subject.

We distinguish between ``default'' conservation of e.g. energy, which is conserved across the ensemble of pump and harmonic light (and medium), and Noether-conserved quantities within the harmonic spectrum. For e.g. ``default'' energy conservation, the energy of the harmonic mode equals the sum of the energies of the photons absorbed in the process. A Noether-conserved quantity within the spectrum has the same value for the pump modes and all generated harmonics. For example, in Ref. \cite{torusknot}, both energy and angular momentum are conserved separately, but the actual values are different for different harmonics (obviously). On the other hand, we find that $\tau\omega_n/\sigma_n - \ell_n/\sigma_n = \gamma$, i.e. a Noether-conserved quantity with the same value for both pump modes and every harmonic mode.

In its default form, the Mathieu equation for a function $f(t)$ is written as $f'' + \omega_1^2[a_1 + a_2 \cos(\omega_2 t)]f = a_3\cos(omega_3 t)$. In the ``default'' use of this equation, the frequencies $\omega_i$ are independent and usually set by outside conditions (e.g. $\omega_2$ may be set by a periodic medium, independent of the pump laser modes). For many cases of beat-wave or harmonic generation, these frequencies are not independent but obey additional resonance conditions, imposed by the pump waves themselves. For harmonic generation by two modes A and B, we have e.g. $\omega_2 = \omega_A/\sigma_A - \omega_B/\sigma_B$ (the pump waves impose their own ``grating'') and either $\omega_3 = \omega_A/\sigma_A$ or $\omega_3 = \omega_B/\sigma_B$, or both. For the scattered wave in stimulated Raman scattering \cite{forslund} or beat-wave processes,, we have $\omega_1 = \omega_2 = \omega_p$ and $\omega_3 = \omega_B/\sigma_B = \omega_A/\sigma_A \pm \omega_p$. For the plasma wave in stimulated Raman scattering, we even have $\omega_1 = \omega_2 = \omega_3 = \omega_p$. In other words, the term $\epsilon \cos(\omega_2 t) f$ is not determined by outside conditions but by the driving laser beams themselves. It is important to keep this in mind while explaining the harmonic spectra of laser beams in isotropic nonlinear media.

There is also a connection between our work and theorems by Bloch and Floquet \cite{bloch, floquet}.
\begin{enumerate}
\item Our work: we recall that any \emph{allowed symmetry} leaves the (periodic) quantity $\Psi_0^* \Psi_0$ invariant. For lasers interacting with a crystal, the (periodic) quantity like $\Psi_0^* \Psi_0 + \Psi_0 \exp(i\varphi) + V(\varphi)$ should remain invariant \cite{alon98, bayku}. Any \emph{allowed harmonic} $\Psi_n$ must then satisfy: $\Psi_n^* \Psi_0$ is invariant under any allowed symmetry.
\item Bloch's theorem: any solution $\Psi_n$ to the time-independent Schr\"odinger equation $-\Psi_{xx} + V(x)\Psi = E\Psi$ with a periodic potential $V(x)$ satisfies: $\Psi_n \exp(-i \vec{k}\cdot \vec{r})$ is periodic in space with the same discrete symmetries as $V(x)$. Thus, the potential dictates the allowed symmetries and Bloch's theorem dictates that allowed solutions must incorporate all allowed symmetries. When we identify the incoming plane wave $\exp(i \vec{k}\cdot \vec{r})$  with $\Psi_0$, the correspondence with our work becomes obvious.
\item Floquet's theorem resembles Bloch's theorem, but deals with functions that are periodic in time rather than space. Let $d\Psi/dt = A(t) \Psi(t)$ where $A(t+T) = A(t)$ for all $t$, and define $f(t) = \Psi^{-1}(0) \Psi(t)$. Then $f(t+T) = f(t) f(T)$ for all $t$ and there exist (possibly complex) matrices $B$ and $P(t)$ such that $f(T) = \exp(TB)$, $P(t+T) = P(t)$ and $f(t) = P(t) \exp(tB)$ for all $t$. When we identify $\exp(tB)$ with $\Psi_0$, the correspondence with our work becomes obvious.
\item In each case, we see that the available symmetries dictate the regular steps of the harmonic progression relative to some starting point (pump wave): $\Psi_0$ in our work, $\exp(i \vec{k}\cdot \vec{r})$ in Bloch's Theorem and $\exp(tB)$ in Floquet's Theorem. The allowed symmetries are then dictated by periodic quantities (``lattices''): $\Psi_0^* \Psi_0$ for harmonic generation, the potential $V(x)$ in the Schr\"odinger equation and the function $A)t)$ in Floquet's theorem. Allowed harmonics must then preserve the following quantities under all allowed symmetries: $\Psi_n \Psi_0^*$, $\Psi_n \exp(-i \vec{k}\cdot \vec{r})$ and $f_n(t) \exp(-tB)$ respectively.
\item Finally, we notice that the ``fundamental modes'' with constant $|\Psi_0|^2$ in our approach correspond to the simplest possible Bloch and/or Floquet states, where the periodic function $\Psi_n \exp(-i K\cdot X)$ is actually constant.
\end{enumerate}
We summarise these results in Table \ref{table:1}.

\begin{table}[h!]
\centering
\begin{tabular}{|c|c|c|c|c|}
\hline
     & Our work & Mathieu & Floquet & Bloch \\
\hline
Translation e.f. & $\Psi_{A,B} = \exp(iK_{A,B} \cdot X)$ & $\exp(\pm ix\sqrt{a})$ & $\exp(tB)$ & $\exp(i\vec{k}\cdot\vec{x})$\\
\hline
Rule for symms. & $\Psi_B^* \Psi_A, \ldots$ inv. & $2q\cos(2x)$ inv. & $A(t+T) = A(t)$ & Periodic pot. $V(x)$ \\
\hline
Rule for $\Psi_n$ & $\Psi_A^* \Psi_n, \Psi_B^* \Psi_n, \ldots$ inv. & $\Psi_n \exp(\mp ix\sqrt{a})$  & $\Psi_n(t) \exp(-tB)$ & $\Psi(x)\exp(-i\vec{k}\cdot\vec{x})$  \\
& under allowed symm. & period $\Delta x = 2\pi$ & has period $T$ & periodic like $V(x)$ \\
\hline
\end{tabular}
\caption{Overview of the correspondence between our work, the Mathieu equation, Floquet and Bloch theory. The first line gives the eigenfunctions for coordinate translations, which form the ``starting point'' for the harmonic progression. The second line gives the term(s) that any allowed symmetry must leave invariant. The third line gives the terms involving any harmonic $\Psi_n$ that must remain invariant under any allowed symmetry, thus shaping the harmonic progression. The similarities between the entries in each row are obvious. }
\label{table:1}
\end{table}


It is worthwhile to note that the generation of novel modes via Eq. (\ref{eq:kprogression}) may be called ``diffraction'' or ``scattering'' for $|\Delta (\omega/\sigma)| = 0$, ``scattering'' for $0 < |\Delta (\omega/\sigma)| < \omega_0$ (e.g. stimulated Raman scattering) and ``harmonic generation'' $ |\Delta (\omega/\sigma)| \geq \omega_0$. Yet in all these cases we are dealing with more or less the same process. (In the case of scattering or diffraction, the range of side bands is often limited by some underlying dispersion relation, making these processes look more different from harmonic generation than they really are.)

There are three potential reasons why the connection between diffraction, scattering and harmonic generation is often obscured. First, scattering usually relies on a grating or crystal lattice that is already present, and thus only needs one fundamental mode. Harmonic generation in a nonlinear medium depends on a ``grating'' provided by the fluctuations in $A^2$, which requires the presence of at least two fundamental modes. Second, a quadratic dispersion relation in $\omega$ usually provides only one branch $\omega > 0$, while a quadratic dispersion relation in $\omega/\sigma$ really provides two branches for $\omega > 0$ and $\sigma = \pm 1$. Diffraction and scattering only need to span one of those branches, while harmonic generation needs to span both because of its large frequency step $\Delta (\omega/\sigma)$, often with two pump modes on opposite branches. Third, a consequence of the previous is that there is no continuous transition from scattering to harmonic generation in most cases, as this is prevented by the dispersion relation.

To summarise this section: we have shown that there are definite connections between (i) Noether's Theorem \cite{noether}, (ii) selection rules for laser harmonic generation in nonlinear media \cite{trines24}, and (iii) the Laue equations for elastic scattering of EM waves by a crystal \cite{laue, kittel}. We have also shown overlap with theorems by Floquet and Bloch \cite{bloch, floquet}. To our knowledge, such a wide range of connections has not been made before. This demonstrates the strength and versatility of our novel approach.

\section{Odd symmetries in coordinate space and tangent space} 
\label{sec:odd}

We start with a mirror symmetry of the coordinates $(t,\vec{x})$. A mirror symmetry in coordinate space imposes a condition on the amplitudes of the coefficients for the modes. For example, a time-reversal symmetry implies that the amplitudes of the complex coefficients for $(\omega/\sigma , \vec{k}/\sigma ) = (+\omega,+\vec{k})$ and $(-\omega,+\vec{k})$ must be the same. The result is a standing wave generated by two counter-propagating CP waves with opposite spin, whose polarisation is locally linear, while the direction of polarisation rotates with a spatial period of $2\pi/|\vec{k}|$: $\Psi \propto \cos(\omega t) \exp(i k_z z)$ or similar.

An ordinary linearly polarised laser pulse, e.g. $\Psi \propto \cos(\omega t - k_z z)$, exhibits a mirror symmetry with respect to the hyperplane $\omega t - k_z z = 0$, which dictates that the coefficients for $(\omega/\sigma , \vec{k}/\sigma ) = (+\omega,+k_z)$ and $(-\omega,-k_z)$ must be the same; note the contrast with the previous case. The argument of the complex coefficients determines the direction of polarisation. In general, we find that a simple mirror symmetry leads to a more complex field configuration, while a simple field configuration leads to a more complex mirror symmetry.

To study generic mirror symmetries, we use $X=(t,\vec{x})$ and $K=(\omega,-\vec{k})$ as before. We consider a mirror symmetry for the plane defined by $K_0 \cdot X = 0$. For any vector $X$, its mirror image is given by $X' \equiv X - 2K_0(X\cdot K_0)/K_0^2$. Then for some $K_A$, $K_A\cdot X' = K_A\cdot X - 2(K_0 \cdot X)(K_A\cdot K_0)/K_0^2$. For any $\Psi$ that obeys this mirror symmetry, i.e. $\Psi(X) = \Psi(X')$, we find that the amplitudes and phases of the Fourier modes $K_A$ and $K'_A \equiv K_A - 2K_0(K_A\cdot K_0)/K_0^2$ must be the same. This condition applies to both the pump modes and the resulting harmonic spectrum. Similarly, the condition that $\Psi(X) = -\Psi(X')$ under a mirror symmetry dictates that amplitudes of $K_A$ and $K'_A$ must be the same but phases must be $\pi$ apart.

A laser configuration that exhibits a mirror symmetry often has some aspect of a standing wave about it, especially in coordinates that are not directly involved in the symmetry. (i) A time-reversal symmetry for two counter-propagating fundamental modes leads to a standing wave in $(t,z)$: $\Psi \propto \cos(\omega t) \exp(i k_z z)$. (ii) A mirror symmetry with respect to the hyperplane $\zeta = \omega t - k_z z = 0$ for two modes with $\ell/\sigma = -1$ leads e.g. to a pulse with radial polarisation: $\Psi =  \cos(\zeta) \exp(i\varphi)$ i.e. a standing wave in $(\zeta,\varphi)$. The harmonics generated by this configuration all share $\ell/\sigma = -1$ and will recombine to radially polarised odd harmonics of the frequency $\omega$, see e.g. De las Heras \emph{et al.} \cite{heras}. (iii) Similar for the ``bicircular'' case from Hickstein \emph{et al.} \cite{hickstein}: $\Psi =  \cos(\zeta) \exp(i k_x x)$ or a standing wave in $(\zeta, x)$. Note the rotating polarisation vector in both this case and the previous one. (iv) A standing wave with linear polarisation exhibits mirror symmetries with respect to the planes $t=0$ and $z=0$ independently: $\Psi =  \cos(\omega t )\cos(k_z z)$; this requires four independent fundamental modes rather than two. (v) A Hermite-Gaussian mode exhibits mirror symmetries with respect to the planes $\zeta=0$ and $x=0$ independently: $\Psi =  \cos(\zeta )\cos(k_x x)$. Again, this requires four fundamental modes. This discussion shows the importance of discussing mirror symmetries in full 4-dimensional spacetime, beyond the work of Ref. \cite{symmetry1}, where only reflections in a single coordinate (time or space) are discussed.

An inversion symmetry of coordinate space will invert multiple coordinates at the same time. For one dimension, this reduces to a simple mirror symmetry; for two dimensions it becomes a rotation over an angle $\pi$; for three dimensions it becomes an ``improper rotation'', where two coordinates are rotated over an angle $\pi$ and the third one is mirrored. This is an extension of the earlier example of time reversal, where he complex coefficients for $(\omega/\sigma , \vec{k}/\sigma ) = (+\omega,+\vec{k})$ and $(-\omega,+\vec{k})$ were required to be the same: now the components of $K$ that correspond to an ``inverted'' coordinate all get a minus sign, while those corresponding to a non-inverted coordinate do not get one.

Next, we look at mirror symmetries of the space of the field vectors, the so-called \emph{tangent space}. We observe that the mode $\Psi = \exp(i\omega t/\sigma )$ is symmetric under time reversal followed by mirroring of the field with respect to the $x$-axis: $\Psi(-t) = \Psi^*(t)$. Similarly, the mode $\Psi = \exp[i(\omega t - k_z z)/\sigma ]$ is invariant under mirroring in the hyperplane $\omega t - k_z z = 0$ followed by mirroring of the field with respect to the $x$-axis.
The mode $\Psi = \exp i(\omega t/\sigma  + \delta)$ is symmetric under time reversal followed by mirroring of the field with respect to the line $E_y/E_x = \tan(\delta)$: $\Psi(-t) = \exp(i\delta)[\Psi(t)\exp(-i\delta)]^*$. If a field $\Psi$ is symmetric under time reversal (or mirroring in e.g. the hyperplane $\omega t - k_z z = 0$, depending on the situation) followed by mirroring of the field with respect to the line $E_y/E_x = \tan(\delta)$, then the fundamental modes underlying this field must all have the same phase $\delta$.

More examples of this kind can be found in Ref. \cite{symmetry1}. In most cases, it helps to decompose the electric field into its fundamental modes, to get a better idea of what is going on.


\section{Examples}
\label{sec:examples}
In the following examples, we use $X = (\delta t, \delta\varphi)$ and $K = [ \Delta (\omega/\sigma), -\Delta (\ell/\sigma)]$

An example with two fundamental modes with $(\omega/\sigma, \ell/\sigma) = (1,1)$ and $(-2,-1)$, taken from Ref. \cite{torusknot}. Symmetries that leave $A^2$ invariant: $(t,\varphi) \to (t + \epsilon\tau, \varphi + \epsilon)$, $\tau = 2/3$ (continuous) and $\varphi \to \varphi + \pi$ (discrete), leading to $X_1 = \epsilon( \tau,1)$ (infinitesimal), $X_2 = (0, \pi)$, $K_1 = \Lambda(1/\tau, 0)$ (divergent, enforcing $n_1=0$ later) and $K_2 = (-2/\tau,+2)$, so $\Delta(\omega/\sigma, -\ell/\sigma) = n_2 K_2 = -n_2(3,-2)$ and $\Delta(\omega/\sigma, -\ell/\sigma) \cdot X_1 = \tau \Delta (\omega/\sigma) - \Delta (\ell/\sigma) = n_1 = 0$, corresponding to the ``conservation of torus-knot angular momentum''. Using the starting point $(\omega/\sigma, +\ell/\sigma) = (1,1)$, we find $(\omega/\sigma, +\ell/\sigma)_n = (1,1) + n(3,2)$, same as Ref.  \cite{torusknot}. In short: one conserved quantity and one periodic quantity, to give a 1-D harmonic progression. Similar configurations can be found in Refs. \cite{moebius, heras}. In all these cases, the data is represented in an $(\omega, \ell)$ spectrum. Converting this to an $(\omega/\sigma, \ell/\sigma)$ spectrum implies that half the data points are rotated around the origin by $180^\circ$, which results in all points ending up on a signle line.

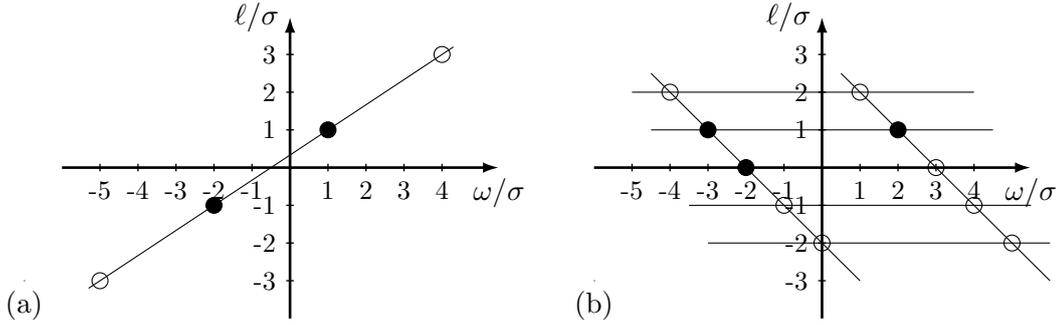
\begin{figure*}[ht]

\begin{tikzpicture}[scale=1]
\begin{scope}[scale=0.5]
    \pgfmathsetmacro{\countA}{2}
    \pgfmathsetmacro{\countB}{2}
    \pgfmathsetmacro{\posXA}{-2}
    \pgfmathsetmacro{\posYA}{-1}
    \pgfmathsetmacro{\incXA}{3}
    \pgfmathsetmacro{\incYA}{2}
    \pgfmathsetmacro{\incXB}{-0.5}
    \pgfmathsetmacro{\incYB}{1.0}
    \pgfmathsetmacro{\hOffsetA}{\incXA - \incYA*\incXB/\incYB}
    \pgfmathsetmacro{\vOffsetB}{\incYB - \incXB*\incYA/\incXA}
    \draw (-7,-3) -- (-7,-3) node[pos=1,below] {(a)};
    \draw[-{Latex[length=6pt, width=4pt]}, line width=1pt] (-6,0) -- (5.5,0) node[pos=1,below] {$\omega/\sigma$};
    \draw[-{Latex[length=6pt, width=4pt]}, line width=1pt] (0,-4) -- (0,4) node[pos=1,left] {$\ell/\sigma$};
    \foreach \x in {-5,...,4} {%
      \ifnum\x=0\relax
      \else
        \draw[line width=0.5pt] (\x,0.1) -- (\x,-0.1) node[below] {\footnotesize \x};
      \fi
    }
    \foreach \y in {-3,...,3} {%
      \ifnum\y=0\relax
      \else
        \draw[line width=0.5pt] (0.1,\y) -- (-0.1,\y) node[left] {\footnotesize \y};
      \fi
    }
    
    \draw[thin] (-5.3, -3.2) -- (4.3, 3.2);
    
    \node[
      draw, circle,      
      minimum size=6pt,  
      inner sep=0pt
    ] at (-5,-3) {};
    \node[
      draw, circle,      
      fill=black,
      minimum size=6pt,  
      inner sep=0pt
    ] at (-2,-1) {};
    \node[
      draw, circle,      
      fill=black,
      minimum size=6pt,  
      inner sep=0pt
    ] at (1,1) {};
    \node[
      draw, circle,      
      minimum size=6pt,  
      inner sep=0pt
    ] at (4,3) {};    
\end{scope}

\begin{scope}[scale=0.5, xshift=14cm]
    \pgfmathsetmacro{\countA}{2}
    \pgfmathsetmacro{\countB}{2}
    \pgfmathsetmacro{\posXA}{2}
    \pgfmathsetmacro{\posYA}{-1}
    \pgfmathsetmacro{\incXA}{-4}
    \pgfmathsetmacro{\incYA}{1}
    \pgfmathsetmacro{\incXB}{-5}
    \pgfmathsetmacro{\incYB}{0}
    \pgfmathsetmacro{\hOffsetA}{\incXA}
    \pgfmathsetmacro{\vOffsetB}{\incYB - \incXB*\incYA/\incXA}
    \draw (-6,-3) -- (-6,-3) node[pos=1,below] {(b)};
    \draw[-{Latex[length=6pt, width=4pt]}, line width=1pt] (-6,0) -- (5.5,0) node[pos=1,below] {$\omega/\sigma$};
    \draw[-{Latex[length=6pt, width=4pt]}, line width=1pt] (0,-4) -- (0,4) node[pos=1,left] {$\ell/\sigma$};
    \foreach \x in {-5,...,4} {%
      \ifnum\x=0\relax
      \else
        \draw[line width=0.5pt] (\x,0.1) -- (\x,-0.1) node[below] {\footnotesize \x};
      \fi
    }
    \foreach \y in {-3,...,3} {%
      \ifnum\y=0\relax
      \else
        \draw[line width=0.5pt] (0.1,\y) -- (-0.1,\y) node[left] {\footnotesize \y};
      \fi
    }

    \foreach \i in {-2,...,2}
        \draw[thin] (-4 + 0.5*\i,-\i) -- (5 + 0.5 * \i, -\i);       
    \foreach \i in {0,1}
        \draw[thin] (-4.5 + 5*\i,2.5) -- (1 + 5*\i, -3);       
    
    \foreach \Point in {(-4,2),(-1,-1),(0,-2),(1,2),(3,0),(4,-1),(5,-2)}
    \node[
      draw, circle,      
      minimum size=6pt,  
      inner sep=0pt
    ] at \Point {};
    \node[
      draw, circle,      
      fill=black,
      minimum size=6pt,  
      inner sep=0pt
    ] at (2,1) {};
    \node[
      draw, circle,      
      fill=black,
      minimum size=6pt,  
      inner sep=0pt
    ] at (-2,0) {};
    \node[
      draw, circle,      
      fill=black,
      minimum size=6pt,  
      inner sep=0pt
    ] at (-3,1) {};
\end{scope}
\end{tikzpicture}

\caption{(a) The $(\omega/\sigma, \ell/\sigma)$ harmonic spectrum based on results by E. Pisanty \emph{et al.} \cite{torusknot}. Unlike in the $(\omega, \ell)$ spectrum, the harmonic peaks (open circles) in this spectrum are all on a single line, while this line and the spacing between peaks are fully determined by the two pump modes (filled circles). The equation for the line, $2(\omega/\sigma)/3 - \ell/\sigma = -1/3 = \mathrm{Const,}$ represents a Noether-conserved quantity, whose value is the same for all harmonics. The positions of the harmonics on the line is given by $(\omega/\sigma-1)/3 = n \in \Z$. (b) The $(\omega/\sigma, \ell/\sigma)$ harmonic spectrum based on results by  G. Lerner \emph{et al.} \cite{symmetry3}. The three independent pump modes (filled circles) are given by $(\omega/\sigma, \ell/\sigma) = (2,1)$, $(-2,0)$ and $(-3,1)$, driving a 2-D harmonic grid. Harmonics (open circles) occur at the crossings of the lines $\ell/\sigma = m\in\Z$ and $(\omega/\sigma + \ell/\sigma +2)/5 = n\in\Z$. }
\label{fig:examples}
\end{figure*}

An example with three fundamental modes, taken from Lerner \emph{et al.} \cite{symmetry3}. This example concerns three modes with (in our notation) $(\omega/\sigma, -\ell/\sigma) = (2,-1)$, $(-2,0)$ and $(-3,-1)$. One discrete symmetry that leaves $A^2$ invariant is given: $X_1 = (2\pi/5)(1, -1)$. The equations given are $\omega/\sigma + \ell/\sigma = 5n - 2$, which describes a collection of lines. To arrive at a collection of points, we need a second symmetry, e.g. $X_2 = (0, 2\pi)$. This symmetry is so obvious that it appears to have been overlooked in Ref. \cite{symmetry3}. However, it is indispensable for a full description of the 2-D harmonic progression. The reciprocal vectors are $K_1 = (5,0)$ and $K_2 = (1,1)$. The harmonic progression is then given by $(\omega/\sigma, -\ell/\sigma)_{nm} = (-2,0) + nK_1 + mK_2$; the above first set of equations by $X_1\cdot(\omega/\sigma, -\ell/\sigma)_{nm} = 2\pi (n-2/5)$ and the second set (needed to arrive at a 2-D grid of points) by $X_2\cdot(\omega/\sigma, -\ell/\sigma)_{nm} = 2\pi m$. Note that we could also have derived $K_{1,2}$ directly from the fundamental modes: $K_1 = (2,-1) - (-3,-1)$ and $K_2 = (-2,0) - (-3,-1)$.

The following case of a 2-D harmonic progression comes up often: when the four fundamental modes are arranged as a parallelogram in 2-D Fourier space. This happens in the case of two laser beams with linear polarisation, but also in the ``perturbed'' case of Pisanty \emph{et al.} \cite{torusknot}. When the opposing sides of the parallelogram exhibit a phase difference of $\pi$, all the coupling terms corresponding to the sides of the parallelogram cancel, and only the terms corresponding to the diagonals are left. While this does not change the positions of the peaks in the 2-D harmonic progression, it will significantly reduce their amplitudes. Also, each peak in the harmonic progression will be associated with a specific ``diagonal'' of the original parallelogram. This happens e.g. when two beams with perpendicular linear polarisations are used and each harmonic is associated with the polarisation of one specific pump beam \cite{long95}, but also in the configuration shown in Fig. 4c of Pisanty \emph{et al.} \cite{torusknot}.

In the following two examples, we look at the interaction of a fundamental laser mode with a crystal with rotational symmetry. We distinguish two types of rotational symmetry for the ``crystal DC mode'' $\Psi_\mathrm{DC}$. Let $\tilde\Psi \equiv \Psi_\mathrm{DC} \exp(-i\varphi)$, then the first type shows $\tilde\Psi(\varphi + 2\pi/N) = \tilde\Psi(\varphi)$ but $\tilde\Psi(\varphi + \pi/N) \not= -\tilde\Psi(\varphi)$ while the second type shows $\tilde\Psi(\varphi + \pi/N) = -\tilde\Psi(\varphi)$. The first type corresponds to DC modes at $\ell/\sigma = nN-1$, $n \in\Z$, while the second type corresponds to DC modes at $\ell/\sigma = (2n+1)N-1$, $n \in\Z$. The configurations of Refs. \cite{alon98, bayku} are of the first type, as are the structured targets used in Ref. \cite{trines24}, while the BBO crystal studied in Ref. \cite{tang} (see below) is of the second type with $N=3$.

An example of a crystal with even symmetry: $\tilde\Psi(\varphi + 2\pi/N) = \tilde\Psi(\varphi)$. Even in its simplest form, there should be at least two discrete symmetries: one provided by the crystal alone and one by the interaction between the crystal and the laser beam. For example, Alon \emph{et al.} \cite{alon98} provide a symmetry $X_1 = (\delta t, \delta\varphi) = (2\pi/N)(1,1)$, which should combine the symmetries of the crystal and the CP driving laser beam. However,  a second symmetry is needed to fully describe the harmonic progression, e.g. the obvious $X_2 = (0, 2\pi)$, which again appears to have been overlooked. This then leads to $K_1 = (N,0)$ and $K_2 = (-1,1)$. The full harmonic progression is then $(\omega/\sigma, -\ell/\sigma)_{mn} = (1 + mN +n, -n )$, $m,n \in \Z$. Harmonics without OAM are given by $\omega/\sigma = 1 + mN$, as in Ref. \cite{alon98}.

An example of harmonic generation in a crystal with odd symmetry: $\tilde\Psi(\varphi + 2\pi/N) = -\tilde\Psi(\varphi)$, given by Lerner \emph{et al.} \cite{symmetry3}, based on work by Tang \emph{et al.} \cite{tang}. This example concerns a mode with $(\omega/\sigma, -\ell/\sigma) = (-1, -\ell_0)$ (the $\vec{e}_+$ mode in this example actually corresponds to $\sigma = -1$ in our notation) hitting a BBO crystal corresponding to $(\omega/\sigma, -\ell/\sigma) = (0,-2)$ and $(0, 4)$ \cite{tang}. The symmetry given is $X_1 = (\pi/3)(\ell_0 + 4, -1)$, and the equations given are (in our notation): $(\omega/\sigma)(\ell_0 + 4) +\ell/\sigma = -6Q+2$. Again this is a collection of lines, so a symmetry must have been overlooked. In this case, it is $X_2 = (2\pi,0)$, ``too obvious''. This leads to $K_1 = (0,-6)$ and $K_2 = (1,\ell_0+4)$, and $(\omega/\sigma, -\ell/\sigma)_{nm} = (-1,-\ell_0) + nK_1 + mK_2$. The above set of equations then reduces to $X_1\cdot(\omega/\sigma, -\ell/\sigma)_{nm} = 2\pi (n-2/3)$, or $[(\ell_0 + 4)(\omega/\sigma) + \ell/\sigma +4]/6 = n \in \Z$, while the second set reads  $X_2\cdot(\omega/\sigma, -\ell/\sigma)_{nm} = 2\pi (m-1)$, or $\omega/\sigma = m-1 \in \Z$. Again, we can calculate $K_{1,2}$ directly from the fundamental modes: $K_1 = (0,-4) - (0,2)$ and $K_2 = (0,4)-(-1,-\ell_0)$.

We note that in many discussions on symmetries in HHG, only one symmetry is provided to describe a harmonic progression in 2-D $(\omega/\sigma, \ell/\sigma)$ space \cite{alon98, torusknot, symmetry3, bayku, li2020, averbukh}. Implicit assumptions will then have to be made to complete the description of the harmonic progression, such as $\ell=0$ \cite{alon98, bayku}, $\Delta (\omega/\sigma) = 1$ \cite{symmetry3, li2020} or $\Delta (\ell/\sigma) = 2$ \cite{torusknot}. A full treatment of these cases including ``overlooked'' symmetries will render such assumptions unnecessary.

An example of an odd coordinate symmetry. A time-reversal symmetry (i.e. the system remains invariant under $t \rightarrow -t$) implies that modes at $\omega/\sigma = \pm \omega_0$ must have the same amplitude and phase, resulting in linear polarisation at $\omega_0$. Time-reversal anti-symmetry ($t\rightarrow -t$ implies $\Psi\rightarrow -\Psi$): modes at $ \omega/\sigma = \pm \omega_0$ must have the same amplitude but opposite phase. Similar for symmetry under reflection in a spatial coordinate, e.g. $x \rightarrow -x$. Symmetry under time reversal plus reflection of a field component (e.g. $t \rightarrow -t$ and $E_x \rightarrow -E_x$ ): all fundamental modes must have the same phase, which then determines the symmetry plane.

Finally, we consider a non-periodic function $f$, and impose the condition that $H = f(z-vt)$ must remain invariant. This gives $(dt,dz) = \epsilon(1,v) + \Lambda(v,-1)$, so $(d\omega,-dk_z) = \epsilon(v,-1) + \Lambda(1,v)$, or $d\omega = v dk_z$. If the EM wave group speed satisfies $d\omega/dk = v$ for some combination $(\omega(k),k)$, then a continuous 1-D ``harmonic progression'' will occur around that point.


\section{Comparison with existing symmetry theory for laser harmonic generation}

To make the connection with previous work exploring the role of symmetries in harmonic generation, we compare our approach to that of Tzur \emph{et al.} \cite{symmetry4}, in particular on the rule for the $C_{N,M}$ symmetry and the new rule when this symmetry is broken by an LP wave with frequency $s\omega$. The $C_{N,M}$ symmetry usually originates from a superposition of two fundamental modes with signed frequencies $\omega/\sigma = M\omega$ and $\omega/\sigma  = (M-N)\omega$. The first fundamental mode remains invariant under any symmetry of the form $\tau_\delta \circ R_{M\delta}$. Under such a symmetry, the second mode may not remain invariant, but it will remain invariant under $\tau_\delta \circ R_{M\delta} \circ \zeta_1$, where $\zeta_1 = R_{-N\delta}$. If we then choose $\omega\delta = 2\pi/N$, we get $\zeta_1 = 1$ and we can pretend it does not exist. For clarity, we choose to keep it around though. The harmonic selection rule then becomes $\omega/\sigma  = M + qN$, which corresponds to repeated application of $\zeta_1$, i.e. one application of $\zeta_1$ for each harmonic step.

Next, we perturb this configuration using an LP wave with frequency $s\omega$. This introduces two new fundamental modes at $\omega/\sigma = \pm s\omega$, which are invariant under the symmetries $\tau_\delta \circ R_{M\delta} \circ \zeta_\pm$ where $\zeta_\pm = R_{(-M \pm s)\delta}$. According to Ref. \cite{symmetry4}, the corresponding selection rule is $\omega/\sigma = M + qN + (a-c)(M+s) + (d-b)(M-s)$, which clearly corresponds to repeated application of $\zeta_1$ and $\zeta_\pm$. We note the similarities to our expression \ref{eq:kprogression}, extended to four fundamental modes. We find that (i) even the system with two fundamental modes exhibiting $C_{N,M}$ symmetry is already ``broken'' compared to a system with only one mode, exhibiting much higher symmetry, (ii) symmetry breaking can be well explained by the addition of more fundamental modes and the increase in interactions between those modes, (iii) the harmonic progression and its dimensionality follow directly from the configuration of fundamental modes and their interactions, even without the need for complex symmetry considerations.

Finally, we note that a rotational symmetry in the $xy$ plane leads to a harmonic progression in the longitudinal OAM. Let $\zeta = z - ct$, then a rotational symmetry in the $\zeta t$ plane may lead to a harmonic progression in the transverse OAM, yielding harmonic modes with so-called ``spatiotemporal vortices'', see e.g. Refs. \cite{transoam1,transoam2,transoam3}. 

There have been several papers on harmonic generation and symmetries in recent years, written by one group \cite{fleischer14, symmetry1, symmetry3, symmetry4, symmetry5}. When comparing these works to our own results, we notice the following.
\begin{enumerate}
\item The authors do not make it clear whether they should be studying symmetries of the nonlinear polarisation $\vec{P}$, the electric field $\vec{E}$ or the hamiltonial $H \propto E^2$, and end up with $\vec{E}$. This should have been the Hamiltonian $H$.
\item Does a fundamental mode have elliptic or circular polarisation? This depends on the specific form of $H$; often one uses $H \propto E^2$, so a fundamental mode should be CP with constant $E^2$. This is especially true for their gas jet experiments (isotropic medium) \cite{fleischer14}, where EP is not fundamental. Compare the results of Refs. \cite{pisanty14, milo15}, who do use CP fundamental modes to analyse the results of Ref. \cite{fleischer14}.
\item The ``microscopic'' symmetry $\gamma$ as defined by the authors should be an isometry; in particular, $\gamma$ should obey $|\gamma r_i - \gamma r_j| = |r_i - r_j|$ for it to commute with their Hamiltonian $H$ for the interaction of light with molecules. For the ``elliptic'' symmetry $e_{n,m}$ this is not the case, so this symmetry should not be employed in this context.
\item In Ref. \cite{symmetry3}, symmetries of the form $(t,x) \to (t,x) + dX$ are being discussed, and rules of the form $(\Delta K) \cdot dX = 2\pi n$ are given for each symmetry. However, one needs two symmetry rules to fully characterise the harmonic progression in 2-D Fourier space, as discussed in Section \ref{sec:examples}, while only one rule is discussed for each example in Ref. \cite{symmetry3}. A rule like $\omega_q/\omega_0 + k_q/k_0 = N$ gives a line or a sequence of lines, not a sequence or lattice of points. The main cause appears to be that symmetries like $t \to t + 2\pi/\omega_0$ or $\varphi \to \varphi + 2\pi$ are so obvious that they are overlooked, see also Ref. \cite{alon98}. We also note that the number of dimensions of any harmonic progression can only be determined if one knows either all the symmetries or all the fundamental modes underlying this progression. Knowing just one symmetry in a multi-dimensional progression is not enough.
\item The octagonal ``quasi-crystal'' of Ref. \cite{symmetry3} can be described using 2 real and 2 synthetic dimensions, no need for 8 dimensions.
\item In recent work, Lerner \emph{et al.} \cite{symmetry5} use $4n$ synthetic dimensions for $n$ laser pulses. Given that a laser pulse with linear polarisation can be described using just 2 fundamental modes, while in most cases these two modes are not fully independent but add just one dimension to the harmonic progression rather than 2, the use of $4n$ dimensions appears superfluous. Using $n$ dimensions for $n$ laser pulses should be enough.
\end{enumerate}

\section{Conclusions}

In this work, we have achieved the following. We have straightened out the definitions and foundations underlying our approach to harmonic progressions, and provide the arguments behind certain choices we make. This complements the results presented in our earlier work \cite{trines24}.
We have developed a generalised description of harmonic generation in a medium with a third-order or second-order nonlinearity (the most common kind). This description unites descriptions based on fundamental modes and/or photon counting with descriptions based on symmetries of the nonlinear model.
We have derived explicit equations describing harmonic progressions in terms of either fundamental modes or symmetries. This includes equations that describe the number of photons needed from each fundamental mode (``channel'') for the lowest-order contribution to each harmonic mode.
We have identified connections between our own model for harmonic progression (\ref{eq:kprogression}), a relaxation of Noether's Theorem \cite{noether} and the Laue equations for elastic scattering of light off crystal lattices \cite{laue}.
We have established the connection between our work and spectral theorems by Bloch and Floquet \cite{bloch, floquet}.
We have applied our model to examples from existing literature \cite{symmetry3} and shown how we can achieve those results in a more streamlined and complete fashion.
We have compared our results to previous work on symmetries in harmonic generation \cite{fleischer14,symmetry1, symmetry3, symmetry4, symmetry5}, and identified what our work offers over and above those previous results.

We foresee the following applications of our work. (i) A standardised algorithm to predict the harmonic progression for a given laser-target configuration.
(ii) For a desired harmonic spectrum, an algorithm to find a laser-target configuration that returns that spectrum.
(iii) Use symmetries to study the interaction of complex light with chiral media.
(iv) The application of known theory of light scattering off crystals to the modelling of harmonic generation.

\end{document}